\documentclass[aps,graphicx,showpacs,amsmath,amssymb]{revtex4}
\usepackage{graphicx}

\begin{document}

\title{Effective error-suppression scheme for reversible quantum computer}

\author{Zhe-Xuan Gong}\email{gongzhexuan@gmail.com}

\affiliation{Department of Physics, Huazhong University of Science and Technology, Wuhan, 430074, China}

\date{\today}

\begin{abstract}
We construct a new error-suppression scheme that makes use of the adjoint of reversible quantum algorithms. For decoherence induced errors such as
depolarization, it is presented that provided the depolarization error probability is less than 1, our scheme can exponentially reduce the final
output error rate to zero using a number of cycles, and the output state can be coherently sent to another stage of quantum computation process.
Besides, experimental set-ups via optical approach have been proposed using Grover's search algorithm as an example. Some further discussion on the
benefits and limitations of the scheme is given in the end.
\end{abstract}

\pacs{03.67.Pp, 03.67.Lx, 89.20.Ff}

\maketitle

\section{Introduction}
The goal of doing quantum computation and quantum information processing reliably in the presence of noise and decoherence has been pursued since the
advent of quantum error correction, which was independently discovered by Shor \cite{Shor95} and Steane \cite{Ste96a}. Later on, several different
approaches to this goal have been studied. Error-avoiding codes \cite{Zan97} depend on existence of subspaces free of decoherence due to special
symmetry properties, and bang-bang type control strategies \cite{Lord98}, including the recent protocol using super-zeno effect \cite{Grover06},
achieve the suppression of decoherence by suitably coupling the system strongly to an external system for short intervals.

Hosten, et al, in their recent paper \cite{Hosten06}, proposed a novel protocol for counterfactual computation using chained quantum zeno effect.
They showed that in certain circumstances, their protocol could also eliminate errors induced by decoherence. However, Mitchison and Jozsa
\cite{Jozsa06} argued that the actual benefit of this protocol seemed quite limited, in that one could resort to much simpler procedure of just
running the computer for many times, which might even eliminate the errors more effectively in most situations. Reasonable as it is, Hosten, et al
\cite{Hosten06'} then pointed out a key benefit of their protocol that truly outruns its rival. Based on their view, one of the potentially important
aspects of any quantum computing protocol involves sending the output \textit{coherently} to another stage of a quantum computer. The simple method
of running the computer for many times cannot output an extremely pure answer easily, because it needs some sort of \textit{majority voting}
\cite{Nilsen} schemes to yield the final answer, whereas the protocol using counterfactual quantum computation can make this benefit by cycling a
single photon many times before sending it to the next processing stage with a low error probability.

Their interesting discussion therefore enlightens one to have a try of combining the profits of both protocols: the simplicity and efficiency of
repeatedly running the computer and the coherent state transmission characteristic of error-suppression protocol with counterfactual computation.
Here we propose a new error suppression scheme that may achieve this nirvana. We noted that the error suppression protocol introduced in
\cite{Hosten06} made use of the adjoint of Grover's search algorithm, which could undo the search process. Unlike their classical counterparts, many
quantum computation processes are unitary, since quantum circuits are fundamentally reversible. One would then ask, naturally, that is it possible to
take the advantage of the reversibility of quantum computers to help fighting against errors? The answer is yes.

\section{A first look at the simplest case}
First we'd like to show the big picture of our scheme in the simplest case. Consider the Grover's search algorithm (GSA) \cite{Grover97} for two
database elements, which is apparently a unitary algorithm if there's no error:
\begin{equation}
|0\rangle \xrightarrow{GSA} |x\rangle\qquad   |x\rangle \xrightarrow{GSA^\dag}|0\rangle
\end{equation}
where $x\in\{0,1\}$ is the marked element. In our theoretical model, decoherence causes depolarization. Assume that, with probability $p\in[0,1]$,
the search algorithm becomes entangled with the environment, and outputs a mixed state:
\begin{equation}
|0\rangle \langle 0| \xrightarrow{GSA}(1-p)|x\rangle \langle x|+p\frac{I}{2}=(1-\frac{p}{2})|x\rangle \langle x|+\frac{p}{2}|1-x\rangle \langle 1-x|
\end{equation}
\begin{equation}
|x\rangle \langle x| \xrightarrow{GSA^\dag}(1-\frac{p}{2})|0\rangle \langle 0|+\frac{p}{2}|1\rangle \langle 1|
\end{equation}
\begin{equation}
|1-x\rangle \langle 1-x| \xrightarrow{GSA^\dag}\frac{p}{2}|0\rangle \langle 0|+(1-\frac{p}{2})|1\rangle \langle 1|
\end{equation}

As a result, if we first run Grover's search algorithm, and then run the algorithm adjoint, we can get the original state $|0\rangle \langle 0|$ with
a probability: $(1-\frac{p}{2})^2+(\frac{p}{2})^2\in[\frac{1}{2},1]$. Note that the first term $(1-\frac{p}{2})^2$ means that both algorithms are
running correctly, while the second term $(\frac{p}{2})^2$ denotes that both have wrong outputs.

\begin{figure}[here]
\includegraphics[width=0.6\textwidth]{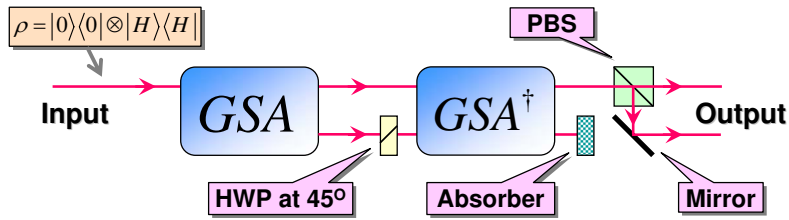}
\caption{\label{fig1} \textbf{Experimental set-up for the simplest reversible quantum algorithm.} The half-wave plate (HWP) at $45^\circ$ rotates the
polarization of photon by $90^\circ$ and polarizing beam splitter (PBS) transmits photon in $|H\rangle$ and reflects $|V\rangle$.}
\end{figure}

We could separate these two terms into orthogonal parts to reduce the depolarization error to $(\frac{p}{2})^2$ by adding an ancillary qubit that
does not enter either algorithm. This will change the process into:
\begin{eqnarray}
|0\rangle \langle 0|\otimes |0\rangle \langle 0| &\xrightarrow{GSA}& (1-\frac{p}{2})|x\rangle \langle x|\otimes |0\rangle \langle
0|+\frac{p}{2}|1-x\rangle \langle 1-x|\otimes |0\rangle \langle 0|\\
&\xrightarrow{CNOT2}& (1-\frac{p}{2})|x\rangle \langle x|\otimes |x\rangle \langle x|+\frac{p}{2}|1-x\rangle \langle 1-x|\otimes |1-x\rangle \langle
1-x|\\ &\xrightarrow{GSA^\dag}& (1-\frac{p}{2})^2|0\rangle \langle 0|\otimes |x\rangle \langle x|+(\frac{p}{2})^2|0\rangle \langle 0|\otimes
|1-x\rangle \langle 1-x| \nonumber
\\&&+(1-\frac{p}{2})\frac{p}{2}|1\rangle \langle 1|\otimes |x\rangle \langle x|+\frac{p}{2}(1-\frac{p}{2})|1\rangle \langle 1|\otimes
|1-x\rangle \langle 1-x|\label{7}\\&\xrightarrow{ABSORB}& (1-\frac{p}{2})^2|0\rangle \langle 0|\otimes |x\rangle \langle x|+(\frac{p}{2})^2|0\rangle
\langle 0|\otimes |1-x\rangle \langle 1-x|\\&\xrightarrow{CNOT1}& (1-\frac{p}{2})^2|x\rangle \langle x|\otimes |x\rangle \langle
x|+(\frac{p}{2})^2|1-x\rangle \langle 1-x|\otimes |1-x\rangle \langle 1-x|\label{9}
\end{eqnarray}
where the operation CNOT1(2) means that the target qubit is 1(2), with the control qubit 2(1), and ABSORB is to terminate the amplitude of both
qubits unless the first qubit is in state $|0\rangle$. This process can be carried out by the simple experimental set-up through optical approach in
Fig. \ref{fig1}, which uses two different paths (upper and lower) as the first qubit of a single photon, and two orthogonal polarization directions
(Horizontal and Vertical) as the ancillary qubit.

\section{Reducing the error rate to zero}
Of courses, we are not to stop at the stage of just reducing the error probability from $p$ to $p^2$. What we aim for is to cut the output error rate
down to an arbitrarily small amount. To achieve it, we \textit{only} need to run $GSA^\dag$ again and again, \textit{i.e.} repeating the process in
Eq.(\ref{7})-(\ref{9}):
\begin{eqnarray}
&&(1-\frac{p}{2})^2|x\rangle \langle x|\otimes |x\rangle \langle x|+(\frac{p}{2})^2|1-x\rangle \langle 1-x|\otimes |1-x\rangle \langle 1-x|\nonumber
\\&\xrightarrow{GSA^\dag}&(1-\frac{p}{2})^3|0\rangle \langle 0|\otimes |x\rangle \langle x|+(\frac{p}{2})^3|0\rangle \langle 0|\otimes |1-x\rangle
\langle 1-x|\nonumber\\&&+(1-\frac{p}{2})^2\frac{p}{2}|1\rangle \langle 1|\otimes |x\rangle \langle x|+(\frac{p}{2})^2(1-\frac{p}{2})|1\rangle
\langle 1|\otimes |1-x\rangle \langle 1-x|\\&\xrightarrow{ABSORB}&(1-\frac{p}{2})^3|0\rangle \langle 0|\otimes |x\rangle \langle
x|+(\frac{p}{2})^3|0\rangle \langle 0|\otimes |1-x\rangle \langle 1-x|\\&\xrightarrow{CNOT1}&(1-\frac{p}{2})^3|x\rangle \langle x|\otimes |x\rangle
\langle x|+(\frac{p}{2})^3|1-x\rangle \langle 1-x|\otimes |1-x\rangle \langle 1-x|\\&\xrightarrow{\ldots\ldots}&(1-\frac{p}{2})^k|x\rangle \langle
x|\otimes |x\rangle \langle x|+(\frac{p}{2})^k|1-x\rangle \langle 1-x|\otimes |1-x\rangle \langle 1-x|
\end{eqnarray}

Accordingly, we find that by running the quantum algorithm (or its adjoint) for a total of k times, we can reduce the probability of getting the
wrong result to $(\frac{p}{2})^k$. And the output error rate $\frac{(\frac{p}{2})^k}{(\frac{p}{2})^k+(1-\frac{p}{2})^k}$ will become near to zero
when $k\rightarrow\infty$ and $p<1$.

\section{General Cases}
Now let's consider a general reversible quantum computer: Suppose this computer has an output register consisted of N qubits that represents the
binary result of the computation, which is initialized to $|0_10_2\ldots0_N\rangle$ at the beginning. We'd also like to assume that the output
register is always in computational basis. (For Grover's search algorithm acting on more than two qubits, one could achieve this by simply replacing
phase inversion operations with phase rotations of angles smaller than $\pi$ \cite{Long01}) Given that without decoherence and noise, the quantum
algorithm (QA) will do the unitary transformation to the output register, and the adjoint algorithm will undo this process: (Here we ignore the extra
qubits the computer will generally require for its input and programming)
\begin{equation}
|0_10_2\ldots0_N\rangle \xrightarrow{QA} |y_1y_2\ldots y_N\rangle\qquad |y_1y_2\ldots y_N\rangle \xrightarrow{QA^\dag} |0_10_2\ldots 0_N\rangle
\end{equation}

When decoherence causes depolarization with probability p, the algorithm will work as:
\begin{equation}
|0_10_2\ldots 0_N\rangle\langle 0_10_2\ldots 0_N| \xrightarrow{QA}(1-\frac{2^N-1}{2^N}p)|y_1y_2\ldots y_N\rangle\langle y_1y_2\ldots
y_N|+\frac{p}{2^N}\sum_{i_1i_2\ldots i_N
\\\neq y_1y_2\ldots y_N}|i_1i_2\ldots i_N\rangle\langle i_1i_2\ldots i_N|
\end{equation}

Applying the error-suppression scheme above with the assistance of an ancillary register consisted of N qubits, we obtain:
\begin{eqnarray}
&&|0_10_2\ldots 0_N\rangle \langle 0_10_2\ldots 0_N|\otimes |0_10_2\ldots 0_N\rangle \langle 0_10_2\ldots 0_N|
\nonumber\\&\xrightarrow{QA,CNOT2}&(1-\frac{2^N-1}{2^N}p)|y_1y_2\ldots y_N\rangle \langle y_1y_2\ldots y_N|\otimes |y_1y_2\ldots y_N\rangle \langle
y_1y_2\ldots y_N|\nonumber\\&&+\frac{p}{2^N}\sum_{i_1i_2\ldots i_N \neq y_1y_2\ldots y_N}|i_1i_2\ldots i_N\rangle \langle i_1i_2\ldots i_N|\otimes
|i_1i_2\ldots i_N\rangle \langle i_1i_2\ldots i_N|\\&\xrightarrow{QA^\dag,ABSORB}&(1-\frac{2^N-1}{2^N}p)^2|0_10_2\ldots 0_N\rangle \langle
0_10_2\ldots 0_N|\otimes |y_1y_2\ldots y_N\rangle \langle y_1y_2\ldots y_N|\nonumber\\&&+(\frac{p}{2^N})^2\sum_{i_1i_2\ldots i_N \neq y_1y_2\ldots
y_N}|0_10_2\ldots 0_N\rangle \langle 0_10_2\ldots 0_N|\otimes |i_1i_2\ldots i_N\rangle \langle i_1i_2\ldots
i_N|\label{17}\\&\xrightarrow{CNOT1}&(1-\frac{2^N-1}{2^N}p)^2|y_1y_2\ldots y_N\rangle \langle y_1y_2\ldots y_N|\otimes |y_1y_2\ldots y_N\rangle
\langle y_1y_2\ldots y_N|\nonumber\\&&+(\frac{p}{2^N})^2\sum_{i_1i_2\ldots i_N \neq y_1y_2\ldots y_N}|i_1i_2\ldots i_N\rangle \langle i_1i_2\ldots
i_N|\otimes |i_1i_2\ldots i_N\rangle \langle i_1i_2\ldots i_N|\label{18}\\&\xrightarrow{\ldots\ldots }&(1-\frac{2^N-1}{2^N}p)^k|y_1y_2\ldots
y_N\rangle \langle y_1y_2\ldots y_N|\otimes |y_1y_2\ldots y_N\rangle \langle y_1y_2\ldots y_N|\nonumber\\&&+(\frac{p}{2^N})^k\sum_{i_1i_2\ldots i_N
\neq y_1y_2\ldots y_N}|i_1i_2\ldots i_N\rangle \langle i_1i_2\ldots i_N|\otimes |i_1i_2\ldots i_N\rangle \langle i_1i_2\ldots i_N|
\end{eqnarray}

Here the operation CNOT1(2) is a group of CNOT gates working respectively on each target qubits in register 1(2) and ABSORB will absorb all the
qubits unless the first register is in the state $|0_10_2\ldots 0_N\rangle$.

The final output error rate, after running the algorithm and its adjoint for k times altogether, can be written as
\begin{equation}
\epsilon(N,p,k)=\frac{(2^N-1)(\frac{p}{2^N})^k}{(2^N-1)(\frac{p}{2^N})^k+(1-\frac{2^N-1}{2^N}p)^k}=\frac{1}{1+\frac{[2^N(\frac{1}{p}-1)+1]^k}{2^N-1}}
\end{equation}

\begin{figure}[here]
\includegraphics[width=0.7\textwidth]{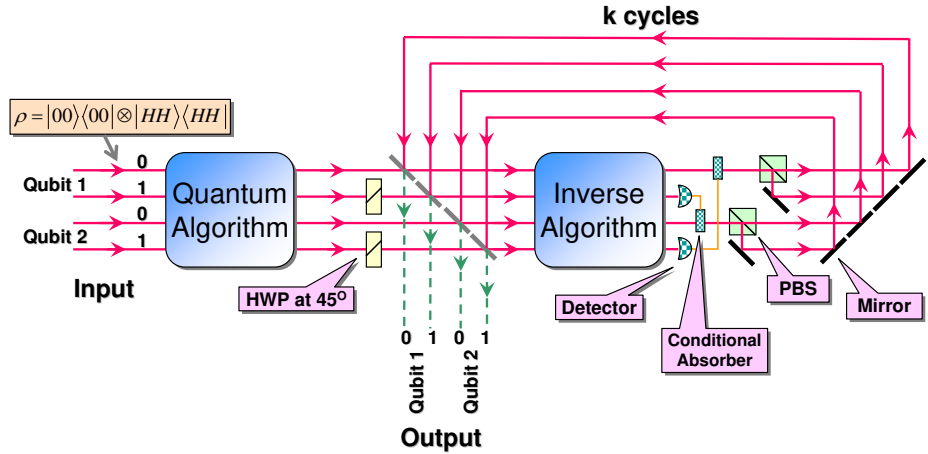}
\caption{\label{fig2} \textbf{Experimental set-up of general error-suppression scheme for reversible quantum computer.} The mirrors in gray color are
inserted after the first cycle and removed right before $k^{th}$ cycles so as to get the final output. The conditional absorber takes effect only if
the detector controlling it has detected a photon.}
\end{figure}

Fig. \ref{fig2} shows a possible set-up for a two-qubit (two-photon) reversible quantum algorithm. It makes use of optical cycles to conveniently
repeat the error-suppression process in Eq.(\ref{17})-(\ref{18}).
\section{Discussions}
\begin{figure}[here]
\includegraphics[width=0.7\textwidth]{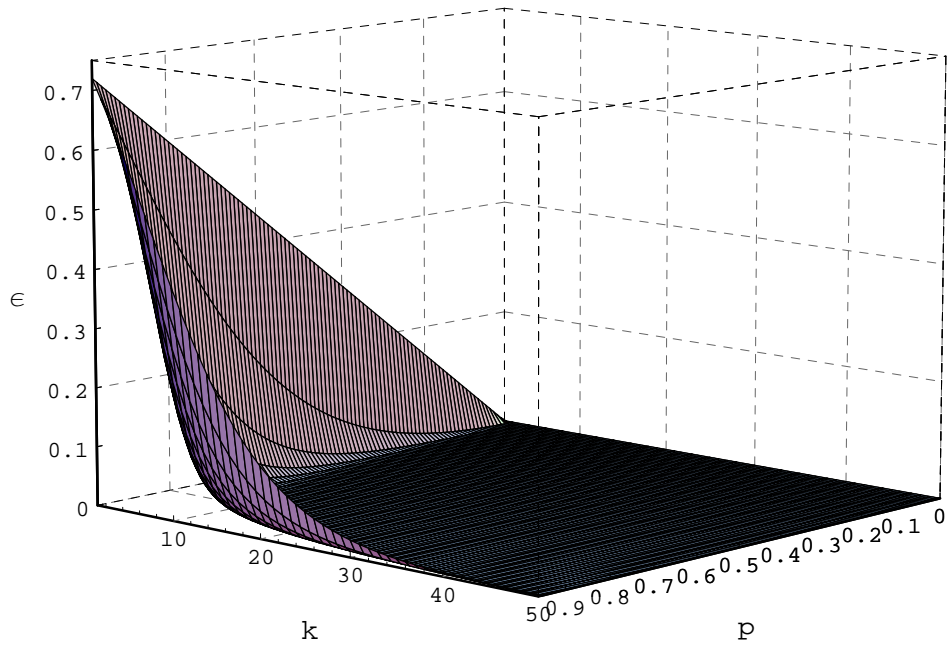}
\caption{\label{fig3} \textbf{Final output error rate function $\epsilon(N,p,k)$ with the case N=2} }
\end{figure}

To check the effectiveness of our error-suppression scheme, we have plotted the function $\epsilon(N,p,k)$ with the case N=2 in Fig. \ref{fig3}. We
can see that as the number of cycles increases, the final output error rate is decreasing to zero exponentially. Even for relatively large p, we only
need to run the quantum computer for a few times to effectively eliminate the errors. (\textit{e.g.} For $N=2, k=10$ and $p=0.5,\epsilon\approx
3\times 10^{-7}$).

\begin{figure}[here]
\includegraphics[width=0.55\textwidth]{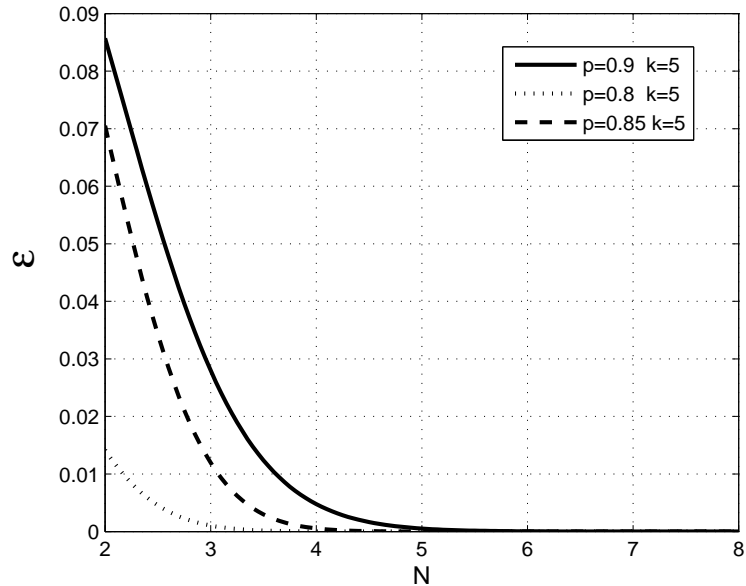}
\caption{\label{fig4} \textbf{Relation between final output error rate $\epsilon$ and the number of qubits $N$}}
\end{figure}

Moreover, the efficiency of our scheme is not compromised by the scale of the reversible computer. Conversely, we show in Fig. \ref{fig4} that when
the number of qubits N increases, the error rate of our final output actually drops down with exponential speed.

\begin{figure}[here]
\includegraphics[width=0.7\textwidth]{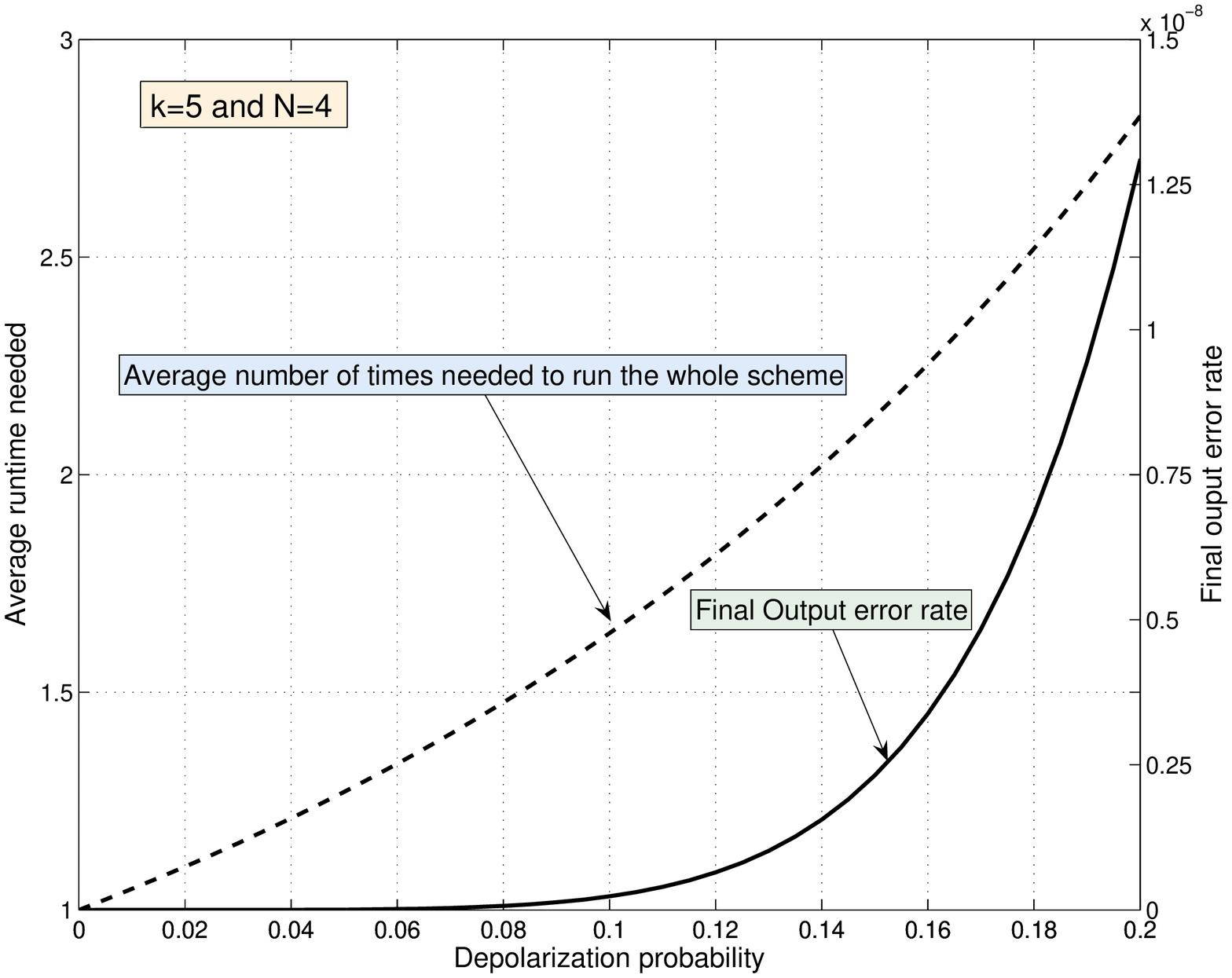}
\caption{\label{fig5} \textbf{Average number of times needed to run the whole scheme / Corresponding final output error rate.}}
\end{figure}

On the other hand, however, it is necessary to mention that with $\epsilon>0$, we always have a probability
$\zeta=1-(\frac{p}{2^N})^k-(1-\frac{2^N-1}{2^N}p)^k$ of failing to obtain a final output, which means that the photons (for optical set-up) are
absorbed during the process of the scheme. Consequently, we have to run the whole algorithm for a second time or more until we obtain a result. (Note
that the majority voting scheme has the same problem). Fortunately, this drawback does not put a high toll on our scheme, since for reasonable values
(relatively small) of p and k,\textit{ e.g.} $p=0.1,k=5,N=4$, we only need to run the whole algorithm 1.6 times on average while the output error
rate is already below $3\times10^{-10}$. Fig. \ref{fig5} gives more in detail.

Another interesting point is that for any $p\in (0,1)$, our error-suppression scheme can give an extremely correct result after enough number of
cycles, but if $p=1$, that is, the quantum computer gives no information in its output (a completely mixed state, $I/2^N$, with the mutual
information being zero), then it is natural to deduce that by whatever means, including our scheme, it is simply impossible to generate any useful
information in the final output. Our scheme acts as an \textit{information amplifier}, but it cannot produce any information from nil.

The process of suppressing errors step by step gradually is analogous to fault tolerant quantum logic using concatenated codes
\cite{Shor96}-\cite{Pres98}. The fault tolerant quantum logic generally consists of three sub processes: encoding, syndrome measurement and recovery,
which might be more complicated to be experimentally carried out compared to our scheme. Additionally, as the threshold theorem for fault-tolerant
computation holds, each component gate of fault-tolerant logic should fail with a probability below the threshold $p_{th}$, the typical value of
which is approximately $10^{-4}$ \cite{Nilsen}. Our scheme puts no limit on the threshold of $p$, except for the extreme case of $p=1$.

There are, nevertheless, a set of limitations for our error suppressing scheme. First, it is best applicable to quantum computers that are
reversible. For quantum algorithms that involve measurement to yield final output, such as Shor's factoring algorithm, we need to take further
measures to make it run in a totally reversible way (This can be achieved in principle, by using some extra registers and unitary operations);
Second, we have to emphasize another premise, that the output of the quantum computer must be in computational basis, \textit{i.e.} it does not allow
superpositions (Note that within the quantum computer, there is no such limit.\textit{e.g.} Grover's search algorithm). It is our hope that this
error-suppression scheme can be of use to a wider scope of quantum computation processes, as well as stimulating further discourse on related topics.

\begin{acknowledgments}
The author would like to thank Professor Richard Jozsa and Professor Robert Griffiths for helpful discussion and suggestions during ICQFT'06 and
AQIS'06, and to thank Professor Ying Wu for kind support and encouragement. The project is supported in part by National Natural Science Foundation
of China under Grant Nos. 10575040 and 90503010, and National Fundamental Research Programme of China under Grant No 2005CB724508.
\end{acknowledgments}


\begin{thebibliography}{99}
\bibitem{Shor95}
P. W. Shor, \textit{Phys. Rev. A} 52, R2493 (1995).
\bibitem{Ste96a}
A. M. Steane, \textit{Phys. Rev. Lett.} 77, 793 (1996).
\bibitem{Zan97}
P. Zanardi and M. Rasetti, \textit{Phys. Rev. Lett.} 79, 3306 (1997).
\bibitem{Lord98}
L. Viola and S. Lloyd, \textit{Phys. Rev. A } 58, 2733 (1998).
\bibitem{Grover06}
D. Dhar, L. K. Grover, and S. M. Roy, \textit{Phys. Rev. Lett.} 96, 100405 (2006)
\bibitem{Hosten06}
O. Hosten, M. T. Rakher, J. T. Barreiro, N. A. Peters and P. G. Kwiat, \textit{Nature}, 439, 949-952 (2006)
\bibitem{Jozsa06}
G. Mitchison and R. Jozsa, \textit{quant-ph}/0606092
\bibitem{Hosten06'}
O. Hosten, M. T. Rakher, J. T. Barreiro, N. A. Peters and P. G. Kwiat, \textit{quant-ph}/0607101
\bibitem{Nilsen}
M. A. Nielsen and I. L. Chuang, \textit{Quantum Computation and Quantum Information} 426-495 (Cambridge Univ. Press, Cambridge, UK, 2000).
\bibitem{Grover97}
L. K. Grover, \textit{Phys. Rev. Lett.} 79, 325-328 (1997).
\bibitem{Long01}
G. L. Long, \textit{Phys. Rev. A } 64, 022307 (2001)
\bibitem{Shor96}
P. W. Shor, \textit{In Proceedings, 37th Annual Symposium on Fundamentals of Computer Science}, pages 56-65, IEEE Press, Los Alamitos, CA, 1996
\bibitem{Pres98}
J. Preskill, \textit{Proc. R. Soc. London A}, 454, 385-410, 1998
\end{thebibliography}
\end{document}